\documentclass[aps,pre,twocolumn,superscriptaddress]{revtex4-2}
\usepackage{amssymb}
\usepackage{amsmath,bm}
\usepackage{graphicx,color}
\usepackage{bbm}
\usepackage[bottom]{footmisc}
\usepackage{multirow}
\usepackage{xcolor}
\usepackage{xpatch} 
\makeatletter   
\usepackage{xpatch} 

\newcommand{\B}[1]{{\bm{#1}}}

\newcommand{\rout}{r_{\rm out}}
\newcommand{\rin}{r_{\rm in}}

\begin{document}
\title{Dynamic Screening by Plasticity in Amorphous Solids}
\author{H. George E. Hentschel}
\affiliation{Dept. of Physics, Emory University, Atlanta Ga. 30322},
\author{Anna Pomyalov} 
\affiliation{Dept. of Chemical Physics, The Weizmann Institute of Science, Rehovot 76100, Israel} 
\author{Itamar Procaccia}
\affiliation{Dept. of Chemical Physics, The Weizmann Institute of Science, Rehovot 76100, Israel}
\affiliation{Sino-Europe Complex Science Center, School of Mathematics, North University of China, Shanxi, Taiyuan 030051, China}.
\author{Oran Szachter}
\affiliation{Racah Institute of Physics, The Hebrew University of Jerusalem, Jerusalem, Israel 9190}
\date{\today}
\begin{abstract}
	In recent work it was shown that elasticity theory can break down in amorphous solids subjected to nonuniform {\em static} loads. The elastic fields are screened by geometric dipoles; these stem from gradients of the quadrupole field associated with plastic responses. Here we study the dynamical responses induced by {\em oscillatory} loads. The required modification to classical elasticity is described. Exact solutions for the displacement field in circular geometry are presented, demonstrating that dipole screening results in essential departures from the expected predictions of classical elasticity theory. Numerical simulations are conducted to validate the theoretical predictions and to delineate their range of validity.  
\end{abstract}
\maketitle
\section{Introduction}
The response of amorphous solids to oscillatory strain is a well studied subject, mainly in protocols employing simple shear, cf. for example \cite{13FFS,20Pri,21RADSM}. While important and indicating a lot of interesting physics, systems under oscillatory shear do not succumb easily to analytic scrutiny, making the comparison of analytic predictions to experiments or simulations quite difficult to accomplish. In this work we turn to oscillatory forcing by another protocol, of an oscillatory inflation of an inner circular boundary, creating non-uniform oscillatory strain on an amorphous solids that is contained between this and a much larger outer circular boundary that is maintained stationary.

The reason for this choice is two-fold. First, in a series of recent works it was shown that classical elasticity theory needs to be reconsidered and amended to describe correctly the mechanical response of amorphous solids subjected to {\em time-independent} non-uniform strains \cite{21LMMPRS,22BMP,22KMPS,22MMPRSZ,23CMP,23JPS,23MMPR}.  A new phase is formed in which the elastic fields are being screened by emergent geometric dipoles in the resulting displacement field. The transition from a quadrupolar screening in the solid phase to a dipolar screening is reminiscent of the {\em structural} transition in 2D crystals from hexagonal solids to hexatics, except that here the transition is in the emergent displacement field and the structure of the solid is always amorphous \cite{23JPS}. 
Our theory provides a classical field theory for describing the mechanical state of a deformed amorphous solids, and it predicts anomalous behavior that is observed in both numerical simulations and experiments. 
In this paper we extend the theory to a new direction, to describe mechanical responses of amorphous solids to {\em dynamical} loading.

The second reason for choosing the present geometry is that it allows considerable analytic progress. We consider the effects of inertia and plastic responses in amorphous matter. For the sake of obtaining analytic solutions we will focus on an amorphous solid contained in an annulus of outer radius $r_{\rm out}$ and inner radius $r_{\rm in}$, such that the inner radius oscillates with a fixed frequency $\Omega$. Initially, before oscillations begin, the material will be brought to mechanical equilibrium in which the resultant force on each particle vanishes. Once oscillations start, we will be interested in the time-dependent  displacement field, evaluated with respect to the initial equilibrated positions. We will show that classical elasticity fails to predict correctly this displacement field, and that an appropriate theory requires taking into account the screening introduced by quadrupolar and dipolar charges that form due to plastic responses. 

The structure of the paper is as follows: in Sect.~\ref{classical} we review the dynamics as expected for our configuration from the solution of the equations of motion dictated by classical elasticity. We then solve analytically for the displacement field of a purely elastic medium which is subjected to oscillatory inflation of an inner boundary. We find that the dynamics is already non-trivial, exhibiting interesting features. Section \ref{screen} introduces quadruple and then dipole screening, leading to predictions of a rich array of expected solutions in which new length scales emerge spontaneously, breaking down elasticity theory. In Sect.~\ref{simulation} we describe simulation results to test the prediction of the theory. The comparison of theory to simulations calls for some careful considerations. First, it is important that the dynamics will describe oscillations around a well defined mechanical equilibrium state. Second, we need to deal with the issue of dissipation. In the numerics we introduce dissipation by damping terms in the particle collisions. In the theory there are dissipative terms proportional the rate of change of displacement fields. This requires careful discussion. Finally, nonlinear effects should be kept at bay. When all these are considered, we find indeed dynamical responses that are in accord with the novel theory. 
Section \ref{discussion} offers conclusion and a discussion of the road ahead.

\section{Dynamics of amorphous solids in two dimensions}
\label{classical}
\subsection{The displacement field in purely elastic solids }

 We start by reviewing the classical approach to dynamics and dissipation \'a la Landau \& Lifshitz \cite{landau1959course}, and then we develop the basic ideas that are called for to accommodate the physics of amorphous solids. 
 
In a purely elastic medium of mass density $\rho$, denote the stress field by $\B \sigma$, and the displacement field by $\B d$. The equation of equilibrium is given by
\begin{equation}
    \nabla \cdot \B \sigma=0 \ .
\end{equation}
i.e. the force per unit volume is zero in equilibrium . Once we are not in mechanical equilibrium the force does not vanish but is equal to the acceleration times the mass per unit volume
\begin{equation}
   \rho \ddot{\B d}=\nabla \cdot \B  \sigma  \ .
   \label{dinsigma}
\end{equation}

For an isotropic body, one can write $\nabla \cdot \B \sigma$ in terms of the displacement to get \cite{landau1959course}
\begin{equation}
	 \rho \ddot{\mathbf{d}}=\mu\nabla^2\mathbf{d}+(\mu+\lambda)\mathbf{\nabla}\left(\nabla\cdot \mathbf{d}\right) \ ,	
    \label{iso}
\end{equation}
where $\mu$ and $\lambda$ are  Lam\'e coefficients.

\subsubsection{Dynamical response to oscillations of an inner circle}
\label{pulsar}
Elasticity theory holds equally well in two and in three dimensions. The theory described below can be easily extended to three dimension, and see for example \cite{23CMP}. In this paper the simulations are performed in two dimensions for computational efficiency, and therefore
we specialize the equations to the case of an oscillating inner circle of initial radius $r_0$ and time dependent radius $r_{\rm in}(t)$, in a system bounded by a fixed, rigid outer circular boundary of radius $r_{\rm out}$.  The simulations were done according to the following protocol. At the beginning of each periodic driving cycle,  the inner circle was inflated to $r_{\rm in}(t=0)=r_0+\Delta$, and the system was equilibrated. Then we periodically inflate and deflate the central disk according to  \begin{equation}
	r_{\rm in}(t)=r_{\rm in}(t=0)-\delta\sin(\Omega t)\,
\end{equation}
where $\Delta, \delta, \Omega$ are  parameters. After a given number of oscillations, the system is again allowed to equilibrate. Then the procedure is repeated for ten times. The reason for this initial set up is to guarantee that the subsequent oscillations are around a well defined equilibrated state and the resulting dynamics is reproducible. It is important that the oscillating displacement field is measured with respect to a well-defined equilibrated state.
In order to compare the numerical  responses to the offered theory, we calculate and analyze the angle-averaged radial displacement.

Clearly, in a realistic amorphous solid such an oscillatory inflation exerts work, and without dissipation the system will not reach a stationary state. Thus in the numerical simulations presented below, cf. Sect.~\ref{simulation}, we will add a small dissipative term to the dynamics of the disks that comprise the amorphous solids. This results in the system reaching an oscillatory steady state.
At this point we continue to focus on an ideal elastic medium to solve Eq.~(\ref{iso}) as it stands, without adding dissipation that in an ideal elastic solid does not exist.
\begin{figure}
	\centering
	\includegraphics[scale=0.6]{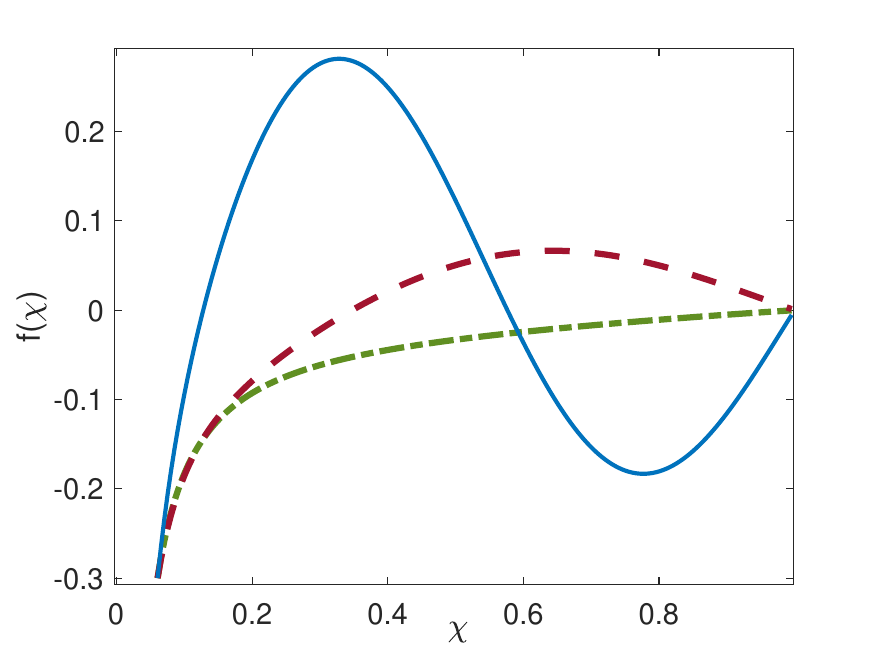}
	\caption{Examples of the function $ f_\omega(\chi)$ according to the purely elastic theory as shown in Eq.~(\ref{bessel3}), for different values of the driving frequency $\omega=1.5,5,7.5$ ( dot-dashed green, dashed red  and solid blue lines, respectively). }
	\label{elastic}
\end{figure}

In the presence of circular symmetry we can focus on the radial component of the displacement field, i.e
\begin{equation}
d_r(r,t) \equiv \B d(r,t) \cdot \B r/r 	 \ .
\end{equation}
The equation for $d_r$ becomes
\begin{equation}
\rho \ddot d_r=\frac{2\mu+\lambda}{r^2} [r^2d_r^{''}+r d_r'-d_r] \ . 
\end{equation}
This equation of motion is solved with the boundary conditions
\begin{eqnarray}\label{bc1}
&&d_r(r_{\rm in},t) =r_{\rm in}(t) -r_{\rm in}(t=0)=-\delta\sin \Omega t\ , \nonumber \\&& d_r(r_{\rm out}) =0 \ .
\end{eqnarray}
The displacement field is measured from the equilibrated configuration obtained after the expansion of the inner boundary to $r_0+\Delta$. Accordingly we seek a solution of the form
\begin{equation}
	d_r(r,t) = f_\Omega(r) \sin (\Omega t) \ , 
\end{equation}
where
\begin{equation}
	 r^2 f_\Omega''(r)+r f_\Omega'(r)+f_\Omega(r) \left(\frac{\rho  r^2 \Omega ^2}{2\mu+\lambda}-1\right) = 0 \ ,
	 \label{bessel1}
	\end{equation}
The dynamic response solves the Bessel equation (\ref{bessel1}). Interestingly, this equation is of the same form as that found in anomalous {\em static} anomalous elasticity cf. Ref.~\cite{21LMMPRS}. Here the frequency related expression acts as dynamical screening. The solution of Eq.~\ref{bessel1} together with boundary conditions \eqref{bc1} in terms of the first order 
Bessel $J_1(r \tilde\omega )$ and Neumann  $Y_1(r \tilde\omega) $ functions is
 \begin{equation}
    f_{\tilde \omega}(r) =- \delta\frac{ \left[Y_1\left(r  \tilde\omega \right) J_1\left(r_{\rm out}  \tilde\omega \right)-J_1\left(r \tilde\omega \right) Y_1\left(r_{\rm out} \tilde\omega \right)\right]}{\left[ Y_1\left(r_{\rm in}\tilde \omega \right) J_1\left(r_{\rm out}\tilde \omega \right)-J_1\left(r_{\rm in} \tilde \omega \right) Y_1\left(r_{\rm out}  \tilde\omega \right)\right]} 	 \label{bessel2}
\end{equation}
where $\tilde \omega \equiv \Omega \sqrt{\rho/(2\mu+\lambda)}$

An important feature of this solution is that $f_{\tilde \omega}$ is fully determined by the geometry of the system, the prescribed frequency, i.e. by $\rin, \rout, \Omega$ and the mechanical properties of the media $\mu$ and $\lambda$. In particular, the number of nodes of the oscillating Bessel functions is  determined by the prescribed frequency $\Omega$. In order to achieve presentation of the results which does not depend on the system size  and material properties, we rewrite the equation and its solutions in dimensionless variables $\chi$ and $\tau$.  This is done using a characteristic length the outer radius $r_{\rm out}$ and characteristic time $t_c\equiv r_{\rm out}/c_d$ where $c_d$ is the dilatational speed 
\begin{equation}
	c_d\equiv \sqrt{(2\mu+\lambda)/\rho} \ .
\end{equation}
With this, 
\begin{equation}
	\chi\equiv r/r_{\rm out}, \quad \tau \equiv t/t_c \ .
\end{equation}
In dimensionless units Eq.~(\ref{bessel2}) reads \begin{equation}
		f_\omega(\chi) =- \delta_\chi \frac{ Y_1\left(\chi  \omega \right) J_1\left(\chi_{\rm out}  \omega \right)-J_1\left(\chi \omega \right) Y_1\left(\chi_{\rm out} \omega \right)}{ Y_1\left(\chi_{\rm in} \omega \right) J_1\left(\chi_{\rm out}\omega \right)-J_1\left(\chi_{\rm in} \omega \right) Y_1\left(\chi_{\rm out}  \omega \right)} 	 \label{bessel3}
	\end{equation}

Note that in this result $\delta_\chi =\delta/ r_{\rm out}$, $\chi_{\rm out}=1$, $\chi_{\rm in}= r_{\rm in}/ r_{\rm out}$, and the frequency $\omega$ is dimensionless.

To get familiarity with the type of functions involved we show in Fig.~\ref{elastic} a few examples of $ f_\omega(\chi)$ as given by Eq.~(\ref{bessel3}). Later we will learn how dissipation and dipole screening change this function in realistic situations.

\subsubsection{Adding dissipation}
\label{adddiss}
In reality, in any experimental or model system, dissipation will be
crucial to balance the work done by an oscillating pulsar at the center of the circular system. In the simulations, we employ disks that interact via Hertzian repulsive forces (see Appendix \ref{Hertz}), but we introduce dissipation by adding to the particle dynamics a dumping force proportional to their velocity $\B v_i$,  $\B F_i\equiv -\tilde \gamma \B v_i$, which is applied to all the disks, to remove any excess energy. In the macroscopic modeling we need therefore to take into account an effective viscous term, that is added to Eq.(\ref{iso}):
\begin{equation}
		\rho \ddot{\mathbf{d}}+\gamma(r)\dot {\B d}=\mu\nabla^2\mathbf{d}+(\mu+\lambda)\mathbf{\nabla}\left(\nabla\cdot \mathbf{d}\right).
	\label{isodiss}
\end{equation}
At this point the function $\gamma(r)$ is kept unspecified, later it will be chosen to accommodate the results of the numerical simulations. For any choice of $\gamma(r)$ the appearance of $\dot {\B d}$ will now mix sine and cosine functions. In dimensionless units we seek a solution for the radial component in the form 
\begin{equation}
	d_\chi(\chi,\tau) =  f_\omega(\chi) \sin (\omega \tau) +g_\omega(\chi) \cos(\omega \tau)\ .
	\label{sum2}
\end{equation}

Next we solve for the functions $f_\omega(\chi)$ and $g_\omega(\chi)$ by substitution the ansatz Eq.~\eqref{sum2} into the boundary value equation Eq.~(\ref{isodiss}), matching together terms in $\cos{\omega \tau}$ and $\sin{\omega \tau}$. We find
\begin{align}
	\label{solution7}
	\begin{split}
	- \omega^2 g_\omega - \gamma(\chi)\frac{t_c}{\rho}\omega~ f_\omega &=\frac{1}{\chi^2}[\chi^2\frac{ d^2 g_\omega}{d\chi^2} +\chi \frac{dg_\omega}{d\chi} - g_\omega] \\
		- \omega^2 f_\omega + \gamma(\chi)\frac{t_c}{\rho}\omega~ g_\omega &= \frac{1}{\chi^2}[r^2\frac{ d^2 f_\omega}{d\chi^2} +\chi \frac{df_\omega}{d\chi} - f_\omega] \ . 
		\end{split}
		\end{align}
These two coupled real equations can be solved by introducing one complex variable 
\begin{equation}
	\label{solution9}
	z(\chi) =  g_\omega(\chi) +i  f_\omega(\chi)  \  .
\end{equation}
Combining Eqs.~(\ref{solution7})  according to Eq.~(\ref{solution9})   we find
\begin{equation}
	\label{solution10}
	-\omega^2 z - i \gamma(\chi)\frac{t_c}{\rho}\omega z = \frac{1}{\chi^2}[\chi^2 \frac{d^2 z}{d\chi^2} +\chi \frac{dz}{d\chi} - z] \ .
\end{equation}

\noindent We can rewrite Eq.~(\ref{solution10}) as
\begin{equation}
	\label{solution11}
\chi^2 \frac{d^2 z}{d\chi^2} +\chi \frac{dz}{d\chi} +z \Big(-1+ \chi^2  [\omega^2  + i \gamma(\chi)\frac{t_c}{\rho}\omega]\Big)  = 0 \ .
\end{equation}

\noindent The solutions of Eq.~(\ref{solution11}) are a combination of a first order 
Bessel $J_1 [\zeta(\chi)]$ and Neumann functions $Y_1[\zeta(\chi)] $ of the complex variable $\zeta(\chi)=\chi~\sqrt{ [ \omega^2  +i \gamma(\chi)\frac{t_c}{\rho}\omega]}$. 

Thus we can write
\begin{equation}
	\label{solution12}
z(\chi)= G J_1( \zeta ) + H Y_1(\zeta )  \ .
\end{equation}
 
To determine the coefficients $G$ and $H$ we need to fit the  boundary conditions. They  are
$f_\omega(\chi_{\rm out})=g_\omega(\chi_{\rm out})=0$ at the outer boundary. At the oscillating inner boundary $f_\omega(\chi_{\rm in}) =-\delta_\chi$, $g_\omega(\chi_{\rm in}) =0$. Accordingly, 
$z(\chi_{\rm in})=-i\delta_\chi$, and $z(\chi_{\rm out})=0$. 
Using these boundary conditions, we define 
\begin{equation}
D = J_1 [\zeta(\chi_{\rm in})] Y_1  [\zeta(\chi_{\rm out})]-J_1 [\zeta(\chi_{\rm out})] Y_1  [\zeta(\chi_{\rm in})] \ .
\label{defD}
\end{equation}
The coefficients in Eq.~(\ref{solution12}) are computed to be
\begin{equation}
	G= -i\delta Y_1  [\zeta(\chi_{\rm out})]/D\, , \quad H = i\delta  J_1  [\zeta(\chi_{\rm out})]/D \ .
\label{defGH}
\end{equation}
Finally, $f_\omega$ and $g_\omega$ of Eq.~(\ref{sum2}) are determined as $\Re(z)$ and $\Im(z)$. 

These exact solutions are expected to 
be relevant as long as dipole screening is absent. To assess the effects of the latter we now add quadrupole and dipole
contributions to the theoretical discussion. 

\section{Anomalous Dynamics due to plastic Screening}
\label{screen}

In this section we derive the equations of motion in the presence of quadrupolar and dipolar screening. This derivation follows the ideas presented in Ref.~\cite{21LMMPRS} for the static problem, supplemented with inertial and dissipative contributions as required. In each case we need to write down the appropriate
Lagrangian which reflects the interactions that are taken into account. We denote the quadrupolar tensor field as $Q^{\alpha\beta}$. Physically, this is the field associated with the eigenstrains of the Eshelby quadrupoles that are formed by plastic responses. The dipolar field $P^\alpha\equiv \partial_\beta Q^{\alpha\beta}$ is simply the gradient of this density. The displacement field is the vector field $d_\alpha$ and the strain tensor
$u_{\alpha\beta} \equiv \tfrac{1}{2} (\partial_\alpha d_\beta + \partial_\beta d_\alpha)$. The Euclidean metric is denoted $g_{\mu\nu}$.

In all cases we start with the 
Euler-Lagrange equations in the standard form  \cite{carroll2019spacetime}, with $\Phi$ playing
the role of a fundamental field, which in our case can be the displacement or the quadrupole density. 
\begin{equation}
	\frac{\partial \mathcal{L}}{\partial \Phi}-\partial_\nu\left(\frac{\partial \mathcal{L}}{\partial\left(\partial_\nu \Phi\right)}\right)=0 \ .
	\label{EL}
\end{equation}
\subsection{Quadrupole Screening}
To derive the equations of motion under quadrupole screening we assume that the density
of quadrupoles is sufficiently low, or that the gradients of their density are negligible.
Thus to quadratic order we can write the Lagrangian for an elastic medium with quadrupoles as
\begin{eqnarray}
    \mathcal{L}&&=\frac{\rho}{2}g^{\mu\nu}\dot{d}_\mu\dot{d}_\nu-\frac{1}{2}\sigma^{\mu\nu}u_{\mu\nu} - \frac{1}{2}\Lambda_{\alpha \beta \gamma \delta}Q^{\alpha\beta}Q^{\gamma\delta}\nonumber \\ &&-\Gamma^{\alpha\beta}_{\gamma\delta}\partial_\alpha d_\beta Q^{\gamma\delta}
\end{eqnarray}
Computing the derivative with respect to the quadrupole field we find

\begin{eqnarray}
    \frac{\partial \mathcal{L}}{\partial Q^{\sigma\kappa}}&=&-\Lambda_{\alpha \beta \gamma \delta}Q^{\alpha\beta}\delta_{\sigma}^{\gamma}\delta_{\kappa}^{\delta}-\Gamma^{\alpha\beta}_{\gamma\delta}\partial_\alpha d_\beta \delta_{\sigma}^{\gamma}\delta_{\kappa}^{\delta}
   \nonumber\\& =&-\Lambda_{\alpha \beta \sigma \kappa}Q^{\alpha\beta}-\Gamma^{\alpha\beta}_{\sigma\kappa}u_{\alpha\beta}=0\ .
   \label{derquad}
\end{eqnarray}
At this point we do not have dipoles in the Lagrangian, or 
\begin{equation}
    \frac{\partial \mathcal{L}}{\partial\left(\partial_\nu Q^{\sigma\kappa}\right)}=0 \ .
\end{equation}
Returning to Eq.~(\ref{derquad}) we rewrite it in the form
\begin{equation}
 Q^{\alpha\beta}=-\Lambda^{\alpha \beta \sigma\kappa}\Gamma^{\gamma\delta}_{\sigma\kappa}u_{\gamma\delta}=\tilde{\Gamma}^{\alpha\beta\gamma\delta}u_{\gamma\delta}
 \label{Qinu}
\end{equation}
Here we denoted the inverse of $\Lambda_{\alpha \beta \sigma \kappa}$ as $\Lambda^{\alpha \beta \sigma\kappa}$ and defined a new (presently unknown) tensor of coefficients $\B {\tilde \Gamma}$.
Eq.~(\ref{Qinu}) is important, showing that the quadrupolar tensor is not some mysterious entity, but that it is induced by the strain field in the system, with the quartic tensor  $\B {\tilde \Gamma}$ providing the link.

Next we consider the derivative with respect to $\partial_\rho d_\sigma$:
\begin{equation}
   \frac{\partial \mathcal{L}}{\partial\left(\partial_\rho d_\sigma\right)}=\rho g^{\mu\nu}\dot{d}_\mu\delta^{\rho}_{t}\delta^{\sigma}_{\nu}-\sigma^{\mu\nu}\delta^{\rho}_{\mu}\delta^{\sigma}_{\nu}-\Gamma^{\alpha\beta}_{\gamma\delta}\delta^{\rho}_{\alpha}\delta^{\sigma}_{\beta} Q^{\gamma\delta}
\end{equation}
Using Eq.~(\ref{EL}) we find
\begin{equation}
  \rho g^{\mu\sigma}\ddot{d}_\mu-\partial_\rho\left( \sigma^{\rho\sigma}+\Gamma^{\rho\sigma}_{\gamma\delta} Q^{\gamma\delta}\right)=0
\end{equation}
At this point we can define a normalized stress field and write
\begin{equation}
	\rho \ddot{d}^\sigma=\partial_\rho \tilde{\sigma}^{\rho\sigma} \ , \quad \tilde{\sigma}^{\rho\sigma}\equiv \sigma^{\rho\sigma} +\Gamma^{\rho\sigma}_{\gamma\delta} Q^{\gamma\delta}= \xi \sigma^{\rho\sigma}  ,
\end{equation}
where the last equality follows from Eq.~(\ref{Qinu}) and homogeneity and isotropy, with $\xi$ being a scalar number. The upshot of this calculation is that the dynamics of the displacement field is unchanged
in form compared to Eq.~(\ref{dinsigma}), except for a renormalization of the elastic moduli. Thus in the present case the introduction of the dissipative term and the analytic solutions for the displacement field follow verbatim the theory presented in the previous section. 

\subsection{Dipole Screening}
The situation changes qualitatively with dipole screening, the equations of motion change their form. To derive these equations we will assert that the renormalization due to quadrupole screening is already included in our Lagrangian, in the form of a renormalized stress tensor (that will be again denoted as $\sigma^{\mu\nu}$). To quadratic order in the fields we therefore write
\begin{eqnarray}
    \mathcal{L}&=&\frac{\rho}{2}g^{\mu\nu}\dot{d}_\mu\dot{d}_\nu-\frac{1}{2}\sigma^{\mu\nu}u_{\mu\nu}-\frac{1}{2}\Lambda_{\alpha \beta}P^\alpha P^\beta-\Gamma^{\alpha}_{\beta}d_\alpha P^\beta\nonumber \\
    &=&\frac{\rho}{2}g^{\mu\nu}\dot{d}_\mu\dot{d}_\nu-\frac{1}{2}\sigma^{\mu\nu} u_{\mu\nu} -\frac{1}{2}\Lambda_{\alpha \beta}\partial_\gamma Q^{\gamma \alpha} \partial_\delta Q^{\delta\beta}\nonumber \\&-&\Gamma^{\alpha}_{\beta}d_\alpha \partial_\gamma Q^{\gamma\beta}\ .
\end{eqnarray}

Computing the derivative with respect to the dipole field we find
\begin{equation}
{\partial_\rho}  \frac{\partial \mathcal{L}}{\partial\left(\partial_\rho Q^{\sigma\kappa}\right)}=-\partial_\sigma[\Lambda_{\alpha \kappa}P^ \alpha+\Gamma^{\alpha}_{\kappa}d_\alpha] =0 \ .
\end{equation}
The expression in the square brackets is a constant, that can be taken as zero using the translational invariance of the displacement field. From this point onward we lose the gauge freedom of the displacement field since we chose a gauge. As before, we use this equation to express the dipole field in terms of the fundamental displacement field:
\begin{equation}
 P^ \alpha=-\Lambda^{\alpha \kappa}\Gamma^{\beta}_{\kappa}d_\beta \ .
\end{equation}

Computing the derivatives with respect to the displacement field and its derivative we find
\begin{equation}
    \frac{\partial \mathcal{L}}{\partial d_\sigma}=-\Gamma^{\alpha}_{\beta} \partial_\gamma Q^{\gamma\beta}\delta^{\sigma}_{\alpha}=-\Gamma^{\alpha}_{\beta} P^\beta\delta^{\sigma}_{\alpha} \ ,
\end{equation}
\begin{equation}
   \frac{\partial \mathcal{L}}{\partial\left(\partial_\rho d_\sigma\right)}=\rho g^{\mu\nu}\dot{d}_\mu\delta^{\rho}_{t}\delta^{\sigma}_{\nu}-\sigma^{\mu\nu}\delta^{\rho}_{\mu}\delta^{\sigma}_{\nu} \ .
\end{equation}
Combining these two equations together we write
\begin{equation}
 -\Gamma^{\sigma}_{\beta} P^\beta- g^{\mu\sigma}\ddot{d}_\mu+\partial_\mu\sigma^{\mu\sigma}=0 \ . \end{equation}
Inverting for $\ddot d^\mu$ we find the modified equation of motion
\begin{equation}
 \ddot{d}^\mu=\partial_\mu\sigma^{\mu\sigma}-\Gamma^{\sigma}_{\beta} \Lambda^{\beta \kappa}\Gamma^{\gamma}_{\kappa}d_\gamma
 \label{broken}
\end{equation}
In a homogeneous and isotropic medium we find
\begin{equation}
\Gamma^{\sigma}_{\beta} \Lambda^{\beta \kappa}\Gamma^{\gamma}_{\kappa} = \kappa^2 g^{\alpha\gamma}\;.
\end{equation}
We note that Eq.~(\ref{broken}) breaks translational symmetry together with the introduction of
a length scale, since $\kappa$ has the dimension of an inverse scale. This is further discussed at
great length (and excuse the pun) in Sect.~\ref{discussion} below. 

\subsection{Solutions of Anomalous Dynamic dipole screening}

As done before, we write $\nabla \cdot \sigma$ in terms of the displacement, to get the explicit equation in the case of dipole screening. As before, we add the dissipative term to the resulting equation and end up with
\begin{equation}
	\rho \ddot{\B d}+\gamma(r)\dot {\B d}=\mu\nabla^2{\bf d}\nonumber +(\mu +\lambda)\nabla(\nabla\cdot{\bf d})+\kappa^2 {\bf d}\,.
	\label{screeneq}
\end{equation}
We note that the addition of the screening term $\kappa^2 \B d$ does not
change the linearity of the equation, and therefore we can use the solution presented in Subsect.~\ref{adddiss} with very little modification. Referring again to a solution in the form of Eq.~\eqref{sum2}, to solve for $f_\omega(\chi)$ and $g_\omega(\chi)$ we only need to change the variable $\zeta$ to a new variable $\tilde \zeta$ where 
\begin{equation}\label{tilde_zeta}
	\tilde \zeta(\chi) \equiv \chi\sqrt {\tilde\kappa^2+\omega^2+i \tilde \gamma(\chi)~\omega} \ ,
\end{equation}
where $\tilde \kappa=\kappa ~t_c/\sqrt{\rho}, \tilde \gamma=\gamma(\chi)~t_c/\rho$.
The Eqs.~(\ref{defD}) and (\ref{defGH}) are unchanged except that the everywhere $\zeta$ is replaced by $\tilde \zeta$. 

\begin{figure}
	\includegraphics[scale=0.6]{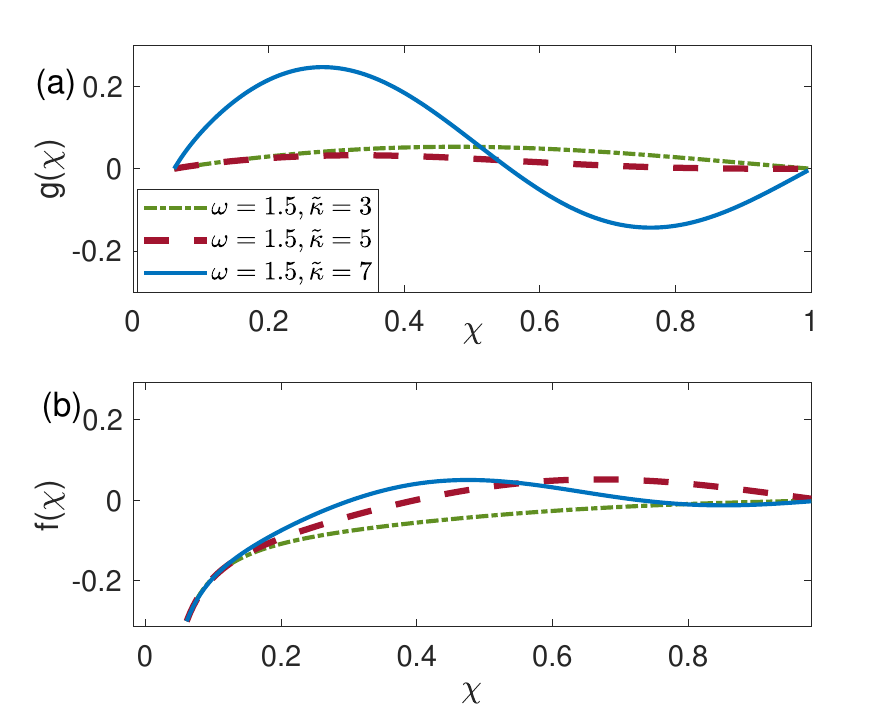}
	\caption{(a) and (b) The functions $g_\omega(\chi)$  and $f_\omega(\chi)$  as predicted with the screening theory, for fixed driving frequency $\omega=1.5$, and fixed dissipation, $\gamma_0=1.0$ and $\epsilon=0.5$,  for three values of the screening parameter $\tilde \kappa=3,5$, and $7$ 
			(dot-dashed green,  dashed red and solid  blue line , respectively. The boundary value $\delta_{\chi}=0.3$.
	}
	\label{screendiss1}
\end{figure}
\begin{figure}
	\includegraphics[scale=0.6]{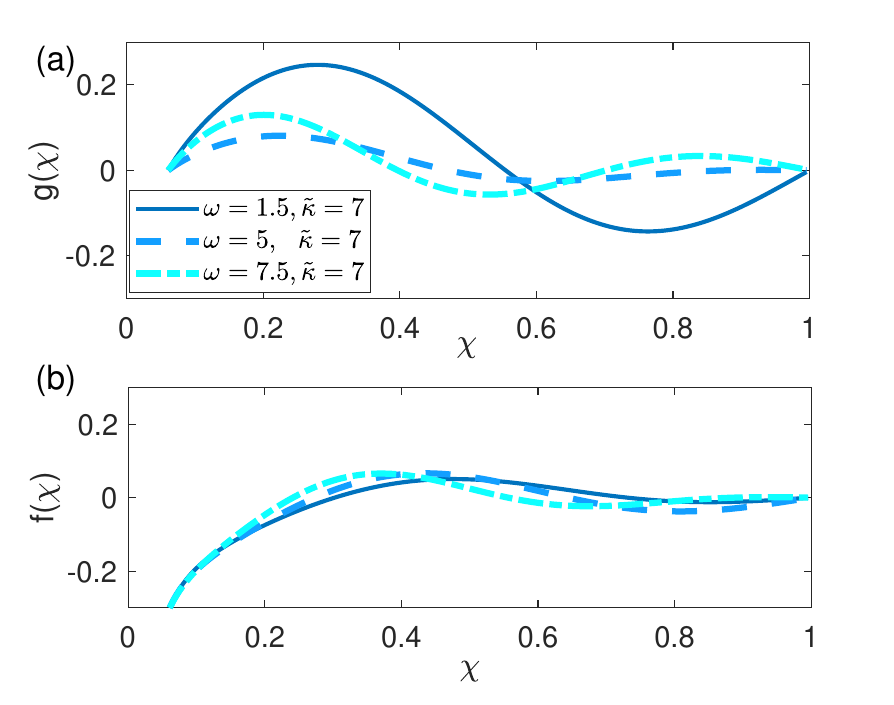}
\caption{ (a) and (b) The functions $g_\omega(\chi)$  and $f_\omega(\chi)$  with the same dissipation parameters as in Fig. \ref{screendiss1} and the  screening parameter $\kappa=7$, for three values of the frequency, $\omega=1.5$ (solid dark blue line-- the same as in Fig. \ref{screendiss1}), $5$ (dashed light blue line), and $ 7.5$ (dot-dashed cyan line).
}
	\label{screendiss2}
\end{figure}

To provide a feeling to the nature of the solutions of Eq.~(\ref{screeneq}) we show in Figs.~\ref{screendiss1} and \ref{screendiss2} the functions $f_\omega(\chi)$ and $g_\omega(\chi)$ for
various frequencies and different values of the screening parameter $\tilde \kappa$. 
 In all cases the dissipation function $\tilde \gamma(\chi)=\gamma_0 [\chi/\chi_{\rm in}(t=0)]^\epsilon$ with  $\gamma_0=1.0$ and $\epsilon=0.5$. The choice of this value of $\epsilon$ has been made purely by comparison to simulations. But one can rationalize it by noting that the geometric spreading of 2-dimensional waves originating from a point source is expected to decay like $r^{-0.5}$ \cite{14FLRS}. On the other hand, since the number of collisions between disks is increasing like $r$, in balance one can expect that a dissipation that increases like $r^{0.5}$ should suffice to keep the energy budget finite. 

\begin{figure}
	\includegraphics[scale=0.6]{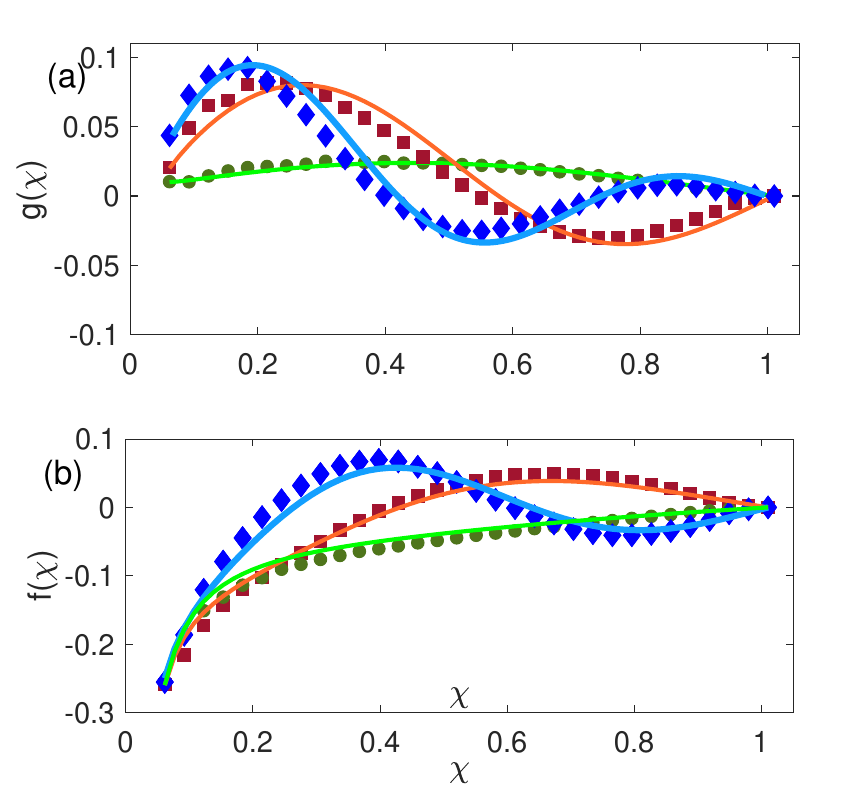}
	\caption{
		(a) Comparison of the measured functions $g_\omega(\chi)$ and $f_\omega(\chi)$ to the predictions of the screening theory, taking account of dissipation. The frequencies are $\omega=1.90$ (green circles), $\omega=4.44$ (red squares) and $\omega=6.34$ (blue diamonds). The respective fitting parameters are $\gamma_0=0.75,\tilde \kappa=1.8$ (green  line),  $\gamma_0=0.9,\kappa=4.5$ (orange  line) and  $\gamma_0=1,\kappa=6.5$  light blue  line). The value of the displacement at the boundary $\delta_1$  differ from the amplitude of the oscillations $\delta_\chi$ due to finite size of the angular-averaging band. For the shown examples the values of  $\delta_1$ and $\delta_2$ (cf. Eq.~(\ref{del12})) correspond to the values of   $f_{\omega}$and   $g_{\omega}$, respectively, at the first available point of $\chi$. }
	\label{compare1}
\end{figure}

\begin{figure}
	\includegraphics[scale=0.6]{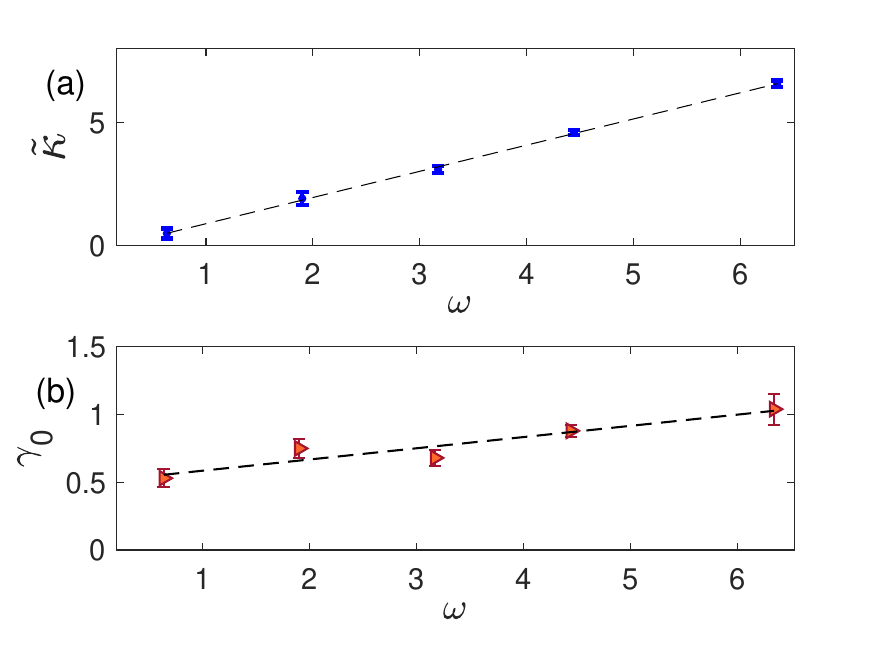}
	\caption{
		The frequency dependence of (a) the anomalous screening parameter $\tilde \kappa$ and  (b) the prefactor of the damping term $\gamma_0$. The values of $\tilde \kappa$ and $\gamma_0$ were averaged over ten different configurations. The error-bars correspond to the standard deviation from the mean values. The lines serve to guide the eye only.}
	\label{f:kappa_gamma}
\end{figure}

\section{Numerical Simulations}
\label{simulation}
\subsection{Setup}
The numerical  setup consists of a 2D concentric circular enclosure
with an inner circle of radius $r_0$, filled with a bi-disperse mixture of frictionless disks. The disks interact through normal Herzian forces only [see Appendix \ref{Hertz}]. The interaction with the outer wall is also Hertzian, with the interaction coefficient equal to the interaction coefficient $K_n$ between disks. Additional viscous dumping force $\B F_i=-\bar\gamma \B v_i$ is applied to all discs to remove excess of energy from the system. Here $\bar\gamma$ is the {\em microscopic} damping coefficient (to be distinguished from the macroscopic $\gamma(r)$ of Eq.~(\ref{isodiss})), and $\B v_i$ is the velocity of the $i$'th disc.
The simulations are carried out using LAMMPS \cite{LAMMPS} for the dynamics of the granular system.  The driving is realized by periodic inflation  of the radius of the inner enclosure wall which is placed at the center of the system. 

The granular system was prepared at a desired packing fraction and pressure,  starting with the
creation of a randomly distributed set of $N=10,000$ particles, half of which has a radius
$r_a = 0.5$ and and the other half with $r_b = 0.7$, with $r_0=4$. Next the system is  equilibrated by energy minimization. Then the radius of the enclosing circle $r_{\rm out}$ is decreased by small decrements from the initial radius to a value corresponding to the largest desired packing fraction $\phi$ according to $r_{\rm out} =\sqrt{(r^2_a+r^2_b)N /2\phi+r_0^2}$. At each step, the system is again equilibrated. A set of configurations for various pressures $p$, or equivalently packing fractions $\phi$, was obtained.

Having a system at mechanical equilibrium we can start oscillating the radius of the inner circle. The procedure followed was explained in Subsect.~\ref{pulsar}. As stated there, to see reproducible results it
is important to guarantee that the oscillations take place around a true
stationary state, in the sense that one is guaranteed that stopping the
oscillations at any point and performing energy minimization would result in the same mechanical equilibrium independent of when and where the oscillatory driving is stopped.  

Another lesson from the numerical experiments is that for the present system of Hertzian disks one is limited in the frequencies of driving that can be simulated. Increasing the frequency too much results in the creation of a hole around the oscillating boundary - the system does not have enough time to return to contact with the boundary when this boundary recedes back too rapidly. We therefore limited the simulations, and the comparison with the theory, to relatively low frequencies as we will see next. It is likely that in systems in which there are attractive forces between the constituents the creation of the hole can be eliminated, and higher frequencies could be studied. On the other hand, it is possible that for any system higher frequencies would trigger nonlinear interactions, requiring additional scrutiny of the equations of motion. At this point we leave these interesting questions to future study.

\subsection{Results for the dominant radial mode} 

To compare the theory with the numerical experiments, we note that the measured functions $g_{\omega}(\chi),f_{\omega}(\chi)$ are calculated by angle-averaging over a band of finite width and therefore we amend the boundary conditions at the inner boundary as
\begin{equation}f_\omega(\chi_{\rm in}) =-\delta_1\ , \quad g_\omega(\chi_{\rm in}) =\delta_2 \ ,
	\label{del12}
	\end{equation} giving
\begin{eqnarray}	\label{defGH1} \nonumber
	G&=& (\delta_2-i\delta_1) Y_1  [\tilde \zeta(\chi_{\rm out})]/D\,  ,\\\  
	H& =&(-\delta_2+ i\delta_1)  J_1  [\tilde \zeta(\chi_{\rm out})]/D\, , \\ \nonumber
	D &=& J_1 [\tilde \zeta(\chi_{\rm in})] Y_1  [\tilde \zeta(\chi_{\rm out})]-J_1 [\tilde\zeta(\chi_{\rm out})] Y_1  [\tilde\zeta(\chi_{\rm in})] \,,
\end{eqnarray}
and $\tilde \zeta(\chi)$ is defined by Eq.\eqref{tilde_zeta}.

In Fig.~\ref{compare1} we present comparisons between the measured functions $g_\omega(\chi)$ and $f_\omega(\chi)$ and the prediction of the theory with dissipation and screening for a number of driving frequencies. The parameters $\tilde \kappa$ and $\gamma_0$ that were used to fit the analytical solution defined by equations \eqref{sum2},\eqref{solution12},\eqref{tilde_zeta}, and \eqref{defGH1}, are typical for our system. The quality of the fits is typical, as long as the frequency of oscillations is not too high, we find very consistent agreement between the analytical theory and the simulations. As can be seen in Appendix B, the transverse components of the displacement field (as well as the nonlinear contributions) are much smaller than the radial ones,
and thus we do not refer to these components in this paper. 

The dependence of the inverse scale $\tilde \kappa$ and the dissipative parameter $\gamma_0$ on the frequency of oscillations appears  linear. In Fig. \ref{f:kappa_gamma} we plot an ensemble-averaged values (over ten independent configuration ) of the $\tilde \kappa$ and $\gamma_0$. The observed linearly is a reflection of the linear regime of the system's dynamics. It is likely that for higher frequencies, with nonlinear effects becoming relevant, also this observed linearity will be lost. 

Our results indicate that as the frequency decreases, $\tilde \kappa$ goes to zero. This means that the low frequency limit does not coincide with a single inflation of the inner circle \cite{21LMMPRS}. The training of our system by repeated oscillation until a stable steady state is obtained removes the anomalous responses to a single inflation step. The system becomes quasi-elastic and not static-anomalous.

\section{Conclusions and the road ahead}
\label{discussion}

It is important to stress that consistent and reproducible results for the measured displacement field under oscillatory forcing depend crucially on insisting upon the existence of true stationary state around which the oscillations take place. As explained, the requirement is that stopping the dynamics at any point in time, and applying energy minimization algorithms result in the same equilibrated configuration up to some severe tolerance requirements. This requires repeated training of the system until such a stable steady state configuration is attained. In terms of the energy picture such a stable steady state corresponds to a sufficiently deep local minimum around which the oscillations take place. 

A second requirement for the successful correspondence between the analytic theory and the simulations is a synchronization between the oscillatory driving and the system's dynamics. In the present case, when the driving frequency exceeds $2\pi$ in dimensionless units, the system reaction cannot follow the driving, entering a dynamical regime that cannot be treated with the present linear theory. Needless to say, this regime is of interest, but is beyond the scope of the present paper.

Finally, the results of this study indicate very strongly that linear classical elasticity fails to provide a valid description of the response of amorphous solids to oscillatory driving, in much the same way as it fails to describe the response to static driving as has been shown in recent work. It is therefore worthwhile to study in the near future also the effect of screening on oscillatory driving by simple or pure shear, since these modes of forcing have attracted a lot of recent interest due to the presence of shear banding and fracture. Such a future direction will automatically necessitate a nonlinear extension of the screening theory, an effort that is under present active research.

\begin{figure}[h!]
	\includegraphics[scale=0.6]{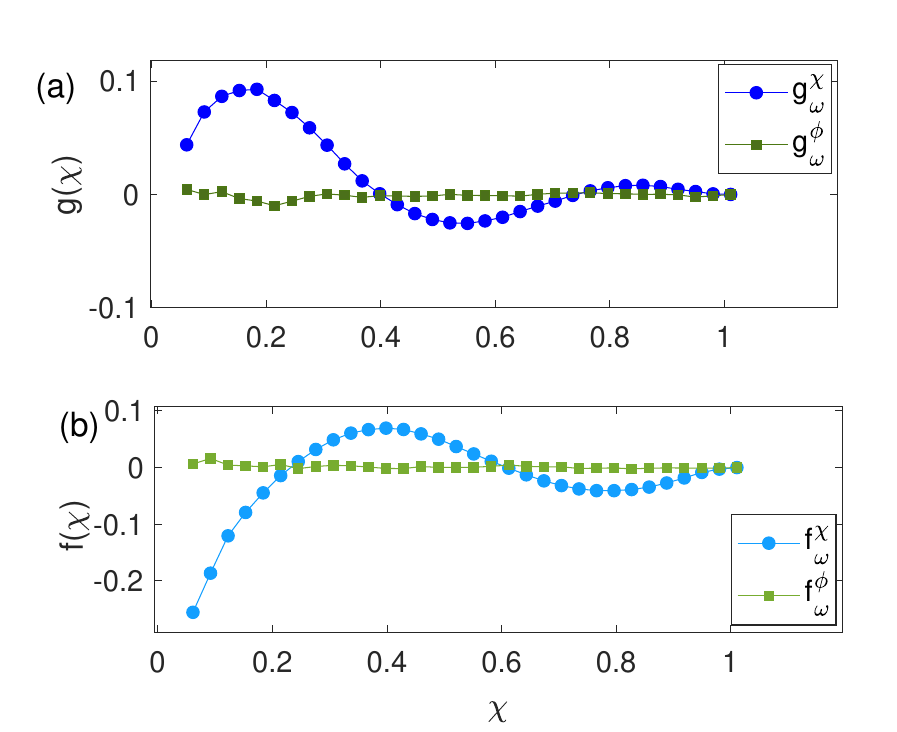}
	\caption{
		The azimuthal contribution to the displacement field (a) $g_{\omega}^{\phi}(\chi)$ and (b)  $f_{\omega}^{\phi}(\chi)$ at the driving frequency (blue circles) are  much smaller that the dominant radial ones $g_{\omega}^{\chi}(\chi)$ and  $f_{\omega}^{\chi}(\chi)$ (green squares). The radial functions are the same is in Fig.\ref{compare1}  with $\omega=6.34$. }
	\label{f:azimuthal}
\end{figure}
\begin{figure}[h!]
	\includegraphics[scale=0.6]{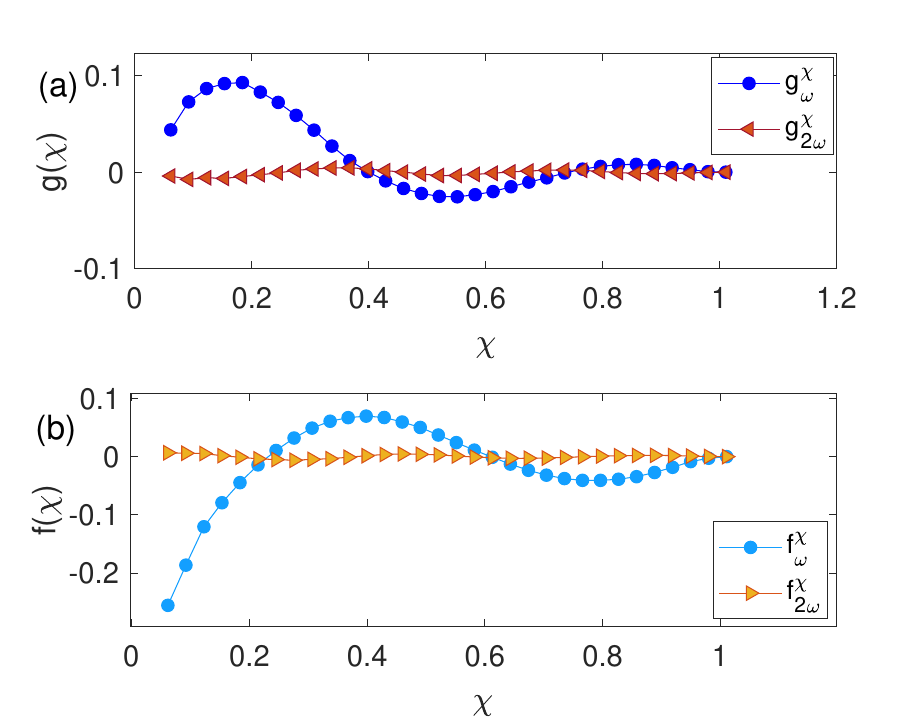}
	\caption{The nonlinear contribution to the displacement field (a)$g_{2\omega}^{\chi}(\chi)$  and (b) $f_{2\omega}^{\chi}(\chi)$ at the double frequency  $2\omega$  (orange triangles) are  much smaller that the dominant radial ones $g_{\omega}^{\chi}(\chi)$ and $f_{\omega}^{\chi}(\chi)$ (blue circles).
		The radial functions are the same as in Fig.\ref{compare1} with $\omega=6.34$. }
	\label{f:nonlinear}
\end{figure}

\begin{appendix}
	\section{Hertzian interaction}
	\label{Hertz}
	The disks interact via normal Hertzian forces
	\begin{equation}
		F=\sqrt{\tilde \delta} \sqrt{\frac{r_a r_b}{r_a+r_b}}[K_n \delta{\bf n}_{ij}-m_{\rm eff} \gamma_n \bf{v}_n]\, .
	\end{equation}
	Here $\tilde \delta$ is the overlap distance of two discs, $K_n$ is the elastic constant of the normal contact, $\gamma_n$ is the viscoelactic constant for normal contact, and $m_{\rm eff}=m_a m_b/(m_a+m_b)$ is the effective mass.
	In our simulations all $m_j=1$, $K_n=2\times 10^4, \gamma_n=500$.
	Additional viscous dumping force $\B F^{v}_i=-\tilde \nu \B v_i$ is applied to all discs to remove excess of energy from the system. Here $\tilde \nu=0.9$ is the damping coefficient and $v_i$ is the velocity of the disc. The simulations
	are carried out using LAMMPS \cite{LAMMPS} for the dynamics of the granular system using LJ units.

		\section{The dominance of the driving mode over nonlinear and azimuthal contributions}
		\label{nonlinear}
		
	Here we present the evidence that the driving mode $g_{\omega}^\chi,f_{\omega}^\chi$ is dominant 	in the considered range of parameters. Since the original displacement field $\B d(\B r)$ has both radial $d_r(r)$ and azimuthal components defined as $\B d^\phi(\B r)\equiv \B d(\B r)-d_r(r) \hat r$. In principle we can solve our equations for both components, but it turns out that the azimuthal component is small compared to the radial one and we did not consider it for the present conditions. In Fig.~\ref{f:azimuthal} we compare the radial contributions at the largest driving  frequency in our range with the azimuthal contribution at the same frequency,  $g_{\omega}^{\phi}(\chi)$ and $f_{\omega}^{\phi}(\chi)$. This justifies the present neglect of this component. In addition, we confirm that the  radial components of the lowest order nonlinear contribution with double frequency, $g_{2\omega}^{\chi}(\chi)$ and  $f_{2\omega}^{\chi}(\chi)$, are also negligible at our range of frequencies, as is shown in Fig.~\ref{f:nonlinear}.

\end{appendix}

\acknowledgements
We thank Michael Moshe for useful discussions at the early stages of this project. This work has been supported by the joint grant between the Israel Science Foundation and the National Science Foundation of China, and by the Minerva Foundation, Munich, Germany. 

\bibliographystyle{unsrt}
\bibliography{bib,ALL.anomalous}
\end{document}